\documentclass[%
 reprint,
 amsmath,amssymb,
 aps,
]{revtex4-2}

\usepackage{graphicx}
\usepackage{dcolumn}
\usepackage{bm}

\begin{document}
\newcommand{\qq}{{\bf q}}
\newcommand{\qqp}{{\bf q}^\prime}
\newcommand{\qqz}{{\bf q}^{\prime\prime}}
\newcommand{\QQ}{{\bf Q}}
\newcommand{\lam}{{\lambda}}
\newcommand{\lamp}{{\lambda^\prime}}
\newcommand{\lamz}{{\lambda^{\prime\prime}}}


\title{First-principles study of UO$_2$ lattice thermal-conductivity: A simple description}

\author{Samira Sheykhi}
\author{Mahmoud Payami}%
 \email{Corresponding Author: mpayami@aeoi.org.ir}
\affiliation{School of Physics \& Accelerators, Nuclear Science and Technology Research Institute, AEOI, 
	P.~O.~Box~14395-836, Tehran, Iran
}%

\date{\today}

\begin{abstract}
	Modeling the high-$T$ paramagnetic state of bulk UO$_2$ by a non-spin-polarized calculation and neglecting 
	the Hubbard-U correction for the $f$ electrons in U atoms, the lattice thermal conductivity of bulk UO$_2$ 
	is investigated by the exact solution of the Boltzmann transport equation for the steady-state phonon 
	distribution function.
	The results show that TA branches corresponding to U-atoms vibrations have the largest lifetimes and 
	therefore have dominant role in thermal conductivity, while the optical branches corresponding mainly to 
	O-atoms vibrations have the shortest lifetimes. 
	Using this simple model, our results for the thermal conductivity show a very good agreement with the experiments. 
	The calculations are repeated for bulk UO$_2$ with different U-235 concentrations of 3\%, 5\%, 7\%, and 
	20\%, and the results show a small decrease of thermal conductivity which arise from scattering of phonons 
	by impurities.
	
\end{abstract}


\maketitle

\section{Introduction}\label{sec1}
One of the vastly used fuels in nuclear power reactors is uranium-dioxide, UO$_2$. The 
performance 
of a nuclear fuel is highly correlated to its thermal conductivity, and therefore studying the 
thermal 
conductivity of nuclear fuel and understanding the mechanisms behind it, is one of the most 
active fields 
of research in nuclear industry.  
Due to the fission processes of the uranium atoms, large amounts of heat are released, leading to 
large 
temperature gradients throughout the fuel rod. Having an efficient thermal conductivity, the 
generated 
temperature-gradients in the fuel are immediately balanced and the heat is easily extracted by 
the coolant 
so that the fuel system does not meet any safety problems because of any temperature-increase or 
thermal 
stresses. 

Experiments have determined the crystal structure of UO$_2$ as a 3k-order antiferromagnet 
(AFM) at 
$T<30^\circ$K, and paramagnetic at higher temperatures \cite{Amoretti,Faber}. The uranium atoms sit on the sites of an 
FCC 
structure with lattice constant $a=5.47\AA$, while the oxygen atoms are positioned at sites with  
$Pa\bar{3}$ symmetry\cite{idiri2004behavior}. Uranium-dioxide is electrically an insulator 
material (the 
so-called Mott insulator), and because of the localized partially-filled $f$-orbitals on U atoms, 
it is a 
strongly-correlated electron system. 
Theoretical description of electronic properties of such a system by ordinary density-functional 
theory 
(DFT)\cite{hohenberg1964,kohn1965self} approximations usually lead to incorrect metallic behavior, and for 
a correct prediction, 
one has to 
somehow take into account the ``localized'' behavior of the $f$ electrons in U atoms. Two methods 
that are 
commonly used for this purpose, are the DFT+U\cite{Dorado09,freyss4} and using orbital-dependent hybrid functional for 
the 
exchange-correlation (XC) part of the energy\cite{SHEYKHI201893}. 
Since in the bulk UO$_2$ there is no delocalized electrons, the thermal properties depends only
on lattice dynamics of the system. The properties of 
lattice 
dynamics are characterized by phonon dispersion, phonon density of states (PhDOS), and anharmonic behaviors.

The lattice dynamics and thermal conductivity of bulk UO$_2$ have been studied by many 
researchers both 
experimentally and 
theoretically. 
For example, Dolling and coworkers\cite{dolling1965crystal}, using the neutron 
inelastic-scattering 
technique, were the first ones that experimentally determined the phonon dispersion and 
density-of-states 
curves of UO$_2$;
Godfrey and coworkers\cite{godfrey1965thermal} have used a radial heat flow technique and 
measured the 
thermal conductivity of polycrystalline UO$_2$ in the range -57$^\circ$ to 1100$^\circ$C;
Goldsmith and coworker\cite{goldsmith1973measurements} using the laser-flash method have measured 
the 
thermal conductivity of porous stoichiometric and hyper-stoichiometric uranium dioxide in the 
temperature 
range of 670-1270$^\circ$K;
Fink and coworkers in two separate works\cite{FINK198117,FINK20001} have reviewed the 
experimental data on 
thermodynamic and transport properties of solid and liquid UO$_2$, and by analyzing the data have 
obtained 
consistent equations for the thermophysical properties.
On the other hand, in theoretical investigations, Motoyama and 
coworkers\cite{motoyama1999thermal} 
resorting to nonequilibrium classical molecular dynamics (MD) have calculated the thermal 
conductivity of 
UO$_2$ pellet; Yamada {\it et al}\cite{YAMADA200010}, employing the partially ionic model within 
their MD 
simulations, have studied the molar specific heat and the thermal conductivity of UO$_2$; Arima 
and 
coworkers\cite{ARIMA200543} using the Born–Mayer–Huggins interatomic potential with the partially 
ionic 
model, have performed equilibrium MD simulations and investigated the thermal properties of 
UO$_2$ and 
PuO$_2$ between 300 and 2000$^\circ$K; Kaur {\it et al}\cite{Kaur_2013} within DFT+U method in generalized 
gradient approximation (GGA)\cite{PBE96} and 
applying an external hydrostatic pressure of 7~GPa, have optimized the geometry of UO$_2$ crystal 
in AFM 1k-order configuration, and then using the resulted lattice parameters have 
studied 
the thermal properties using the density functional perturbation theory (DFPT)\cite{dfpt2001} in conjunction 
with the quasi-harmonic approximation (QHA); Pang and coworkers\cite{pang2013}, in a joint 
experimental and theoretical investigation,    
using high-resolution inelastic neutron scattering, have measured phonon lifetimes and dispersion 
of UO$_2$ at 295 and 1200$^\circ$K, and then analyzing the calculated result of thermal 
conductivity within the relaxation-time approximation (RTA), have concluded that longitudinal 
optical (LO1) branch of phonons carries the 
largest amount of heat; Resnick and coworkers\cite{RESNICK2019} have used an MD simulation to 
study the thermal 
transport of plutonium dioxide and uranium dioxide with point defects; Torres and coworker\cite{TORRES2019137} had performed the solution of BTE on top of 
a DFT+U calculation with 1k-order AFM configuration for UO$_2$, but because of the diversity of their work, they had not provided a satisfactory analysis of the problem.  

One of the ways to investigate the heat transport in solids is using the Boltzmann transport 
equation (BTE). The earlier BTE studies on lattice thermal conductivity of UO$_2$ were based on 
the relaxation-time approximation (RTA). Although the RTA gives good results in cases where the 
``umklapp'' scattering processes are dominant, the exact solution of the BTE is needed whenever the 
``normal'' processes dominate. In this work, we have obtained the 
exact solution of BTE for phonons' distribution from which the lattice thermal conductivity is 
calculated. Our results show a very good agreement with the experiments. 

In nature, uranium is found as U-238 (99.2739\%), U-235 (0.7198\%), and a very small amount of 
U-234 (0.0050\%). Since in different nuclear fuels the relative abundances of U-238 and U-235 are 
different, the scattering rates by ``impurities'' differs from one fuel to other and this affects 
the thermal conductivity of the fuel. We have therefore repeated the procedure for the cases 
U-235 (3\%), (5\%), (7\%), and (20\%), and the results showed a small decrease of thermal 
conductivity by increasing the impurity level.  

The organization of this paper is as follows. In Section~\ref{sec2} we present the computational details; 
in Section~\ref{sec3} the results are presented and discussed; Section~\ref{sec4} concludes this work. 
Finally, in \ref{app} the convergency issues of our calculations are detailed.  

\section{Computational details}\label{sec2}
Since the simulation of the high-$T$ paramagnetic state of UO$_2$ system with randomly oriented 
magnetic moments requires a very 
large supercell, it is computationally very prohibitive and it is common to model it with a 1k-order AFM. 
However, in this work, we model the  
paramagnetic state of bulk UO$_2$ by a non-spin-polarized calculation, and neglect the Hubbard-U 
correction for the localized orbitals in the study of lattice dynamics. It has already been observed that although not using the Hubbard correction in the DFT calculations gives incorrect metallic ground state for UO$_2$, but surprisingly the phonon properties are comparable with experiment \cite{Xiao-Dong13}. 

\subsection{Geometry and Harmonic lattice dynamics}
For the electronic structure calculations, we have used the first-principles DFT method as implemented in 
the Quantum-ESPRESSO code package\cite{giannozzi2009quantum} and used Ultra-soft pseudopotentials 
with the rev-PBE XC-functional\cite{revpbe1998}. The reason for choosing the rev-PBE method is that it gives the best lattice constant for this system. Comparing the results for functionals LDA\cite{kohn1965self}, 
PBE\cite{PBE96}, PW91\cite{PW91}, and rev-PBE; we obtained the values of 5.27, 5.36, 5.36, and 5.39~\AA for the lattice constants, respectively.  
The kinetic-energy cutoffs for the expansion of the Kohn-Sham (KS)\cite{kohn1965self} orbitals and charge 
densities were chosen as 60 and 600~Ry, respectively. For the integrations over the Brillouin-zone (BZ), a  
$\Gamma$-centered $8\times 8 \times 8$ grid were used. The optimization of the geometry was performed with 
a maximum $10^{-5}$Ry/au of residual force on each atom. The convergency tests with respect to the 
parameters are detailed in \ref{app}.   

To calculate the second-order (2nd) interatomic force constants (IFCs) and phonon frequencies, we have 
employed the DFPT method implemented in QE package for the optimized geometry of 3-atom primitive cell 
using a $\Gamma$-centered $12\times 12 \times 12$ uniform grid in the 
reciprocal space (The convergency test is presented in \ref{app}). To ensure the translational invariance 
of the symmetrized dynamical matrix, the acoustic sum-rule was applied. We have obtained the phonon frequencies by diagonalizing the calculated dynamical matrix. The effect of different atomic 
masses of the uranium isotopes on the phonon frequencies was checked and no meaningful differences 
obtained, and therefore, in subsequent calculations all phonon frequencies were calculated with 
the mass of U-238 isotope.

\subsection{Thermal conductivity}
The lattice thermal conductivity of bulk UO$_2$ is calculated by solving the 
linearized-BTE\cite{peierls1997kinetic,ziman1960electrons} for the steady-state phonon distribution 
function $f_\lambda$:
\begin{equation}\label{bte}
\nabla T \cdot {\bf v}_\lambda \frac{\partial f_\lambda}{\partial T} = \left. \frac{\partial f_\lambda}{\partial t}\right|_{\rm scattering},
\end{equation}
where the left hand side of the equation corresponds to the phonon diffusion due to
temperature gradient, and the term in the right hand side is the time rate of change of phonon distribution due to all allowed scattering processes.
Here, ${\bf v}_\lambda$ is the group velocity of phonon in mode $\lambda$ and $\lambda\equiv(s,\qq)$ with  $s$ and $\qq$ being the phonon 
branch index and wave vector in reciprocal space, respectively. 
In the linearized-BTE, the distribution $f_\lambda$ differs from the equilibrium Bose-Einstein 
distribution function $f_\lambda^0=1/(\exp(\beta\hbar\omega_\lambda)-1)$ by a linear term in $\nabla T$.
Here $\beta=1/k_BT$ and $\omega_\lam$ is the phonon frequency in mode $\lambda$. The solution of the BTE 
is done using a full iterative  
algorithm\cite{OMINI1995101,Lindsay2010,Li2012,mingo2014ab} employing ShengBTE code 
package\cite{LI20141747}.
To calculate the thermal conductivity, the third-order (3rd) IFCs were computed up to the second shell of neighbors in a $3\times 3\times 3$ supercell with a $\Gamma$-centered $2\times 2\times 2$ uniform mesh of $k$ points. 
The 3rd-IFCs corresponding to the displacements of $(i,j,k)$ atoms along directions 
$(\alpha,\beta,\gamma)$ were computed using a three-point finite-difference method. To ensure the 
translational invariance, we have imposed the constraint $\sum_i \Phi_{\alpha\beta\gamma}(i,j,k)=0$ 
according to the prescription given by Esfarjani~{\it et al.}\cite{Esfarjani2008} and Li~{\it et al.}\cite{Li2012inv}.

For the scattering contributions, which is treated within perturbation theory, we have considered scattering by isotopes\cite{Tamura83,Tamura84} and all the three-phonon processes satisfying the energy and quasi-momentum conservation:
\begin{eqnarray}
\omega_\lam \pm \omega_\lamp = \omega_\lamz\label{econv}, \\ 
\qq \pm \qqp = \qqz +\QQ\label{qconv},
\end{eqnarray}
where $\QQ$ is a reciprocal lattice vector.
The linearized-BTE may be recasted in a form expressed in terms of a set of coupled equations for the 
phonon lifetimes, $\tau^{(\alpha)}_\lambda$ as\cite{OMINI1995101,Lindsay2010,Li2012,mingo2014ab}:
\begin{equation}\label{tau}
\tau_\lambda^{(\alpha)} = \tau_\lambda^0 (1+\Delta_\lambda^\alpha),
\end{equation}
where $\tau^0_\lambda$, the phonon lifetime in the single-mode relaxation-time approximation (RTA), is defined 
by: 
\begin{equation}\label{tau0}
\frac{1}{\tau_\lambda^0}=\frac{1}{N}\left( \sum_{\lamp\lamz}^+ \Gamma^+_{\lam\lamp\lamz} + \frac{1}{2}\sum_{\lamp\lamz}^- \Gamma^-_{\lam\lamp\lamz} +
\sum_{\lamp} \Gamma_{\lam\lamp}\right),
\end{equation}
in which $N$ is the number of unit cells, and $\tau^{(\alpha)}_\lam$ corresponds to phonon modes propagating in $\alpha$ direction. The ``+'' and ``--'' symbols denote the sums are over two different types (''combination'' and ``decay'' processes, respectively) of three-phonon processes defined by Eq.(\ref{econv}) and Eq.(\ref{qconv}) including both ``normal'' ($\QQ=0$) and ``umklapp'' ($\QQ\neq 0$) processes. The quantities $\Gamma^\pm_{\lam\lamp\lamz}$ and $\Gamma_{\lam\lamp}$ are respectively the three-phonon and phonon-impurity scattering rates. The quantity $\Delta^\alpha_\lambda$ in Eq.(\ref{tau}) is defined by:
\begin{eqnarray}\label{Delta}\nonumber
\Delta_\lambda^\alpha=\frac{1}{N}\left(\sum_{\lamp\lamz}^+ \Gamma^+_{\lam\lamp\lamz} (\xi^\alpha_{\lam\lamz}\tau^{(\alpha)}_\lamz - \xi^\alpha_{\lam\lamp}\tau^{(\alpha)}_\lamp)\right.\\ \nonumber
+ \frac{1}{2}\sum_{\lamp\lamz}^- \Gamma^-_{\lam\lamp\lamz} (\xi^\alpha_{\lam\lamz}\tau^{(\alpha)}_\lamz + \xi^\alpha_{\lam\lamp}\tau^{(\alpha)}_\lamp)\\ 
+ \left.\sum_{\lamp}\Gamma_{\lam\lamp} \xi^\alpha_{\lam\lamp}\tau^{(\alpha)}_\lamp \right)
\end{eqnarray}    
where $\xi^\alpha_{\lam\lamp}=v^\alpha_\lamp\omega_\lamp/v^\alpha_\lam\omega_\lam$. 
The solution of Eq.(\ref{tau}) is achieved by the iterative process of 
$\tau_\lam^{(\alpha),(n)} = \tau_\lam^{0} (1+\Delta_\lam^{\alpha,(n-1)})$ in which 
$\Delta_\lam^{\alpha,(0)}=0$ and $\tau_\lam^{(\alpha),(0)}=\tau_\lam^0$. When the self-consistent $\tau_\lam^{(\alpha)}$ is determined, the thermal conductivity tensor is calculated from:
\begin{equation}\label{kappa}
\kappa^{\alpha\beta}= \frac{1}{V k_{B}T^2}\sum_{\lam}(\hbar\omega_\lam)^2 f^0_\lam(f^0_\lam+1)v^\alpha_\lam v^\beta_\lam \tau^{(\beta)}_\lam.
\end{equation} 

Having 2nd- and 3rd-IFCs at hand, the iteration of Eq.(\ref{tau}) is started by calculating the scattering 
rates $\Gamma^\pm_{\lam\lamp\lamz}$ and $\Gamma_{\lam\lamp}$, and $\tau^0_\lambda$ from Eq.(\ref{tau0}), 
and continued by computing $\Delta^\alpha_{\lam}$ from Eq.(\ref{Delta}). At each step of iteration, the 
conductivity is calculated from Eq.(\ref{kappa}) and the iteration continues until the relative change of 
the conductivity norm $\sqrt{\sum_\alpha\sum_\beta |\kappa^{\alpha\beta}|^2}$ is less than $10^{-5}$.      
To calculate the thermal conductivity form Eq.(\ref{kappa}), we have used an appropriate BZ sampling 
density of $10\times 10\times 10$ and Gaussian smearing of 0.1, which gives almost equal conductivity 
values obtained using $30\times 30\times 30$ and Gaussian smearing of 1.0.     

\section{Results and Discussions}\label{sec3}
\subsection{Geometry and Harmonic lattice dynamics}
Using a non-spin-polarized calculation, we have first fully optimized the geometry of the UO$_2$ primitive 
unit cell with space group 225 and obtained the optimized value of lattice constant for the FCC structure 
as $5.396\AA$. 

\begin{figure}
	\centering
	\includegraphics[width=\linewidth]{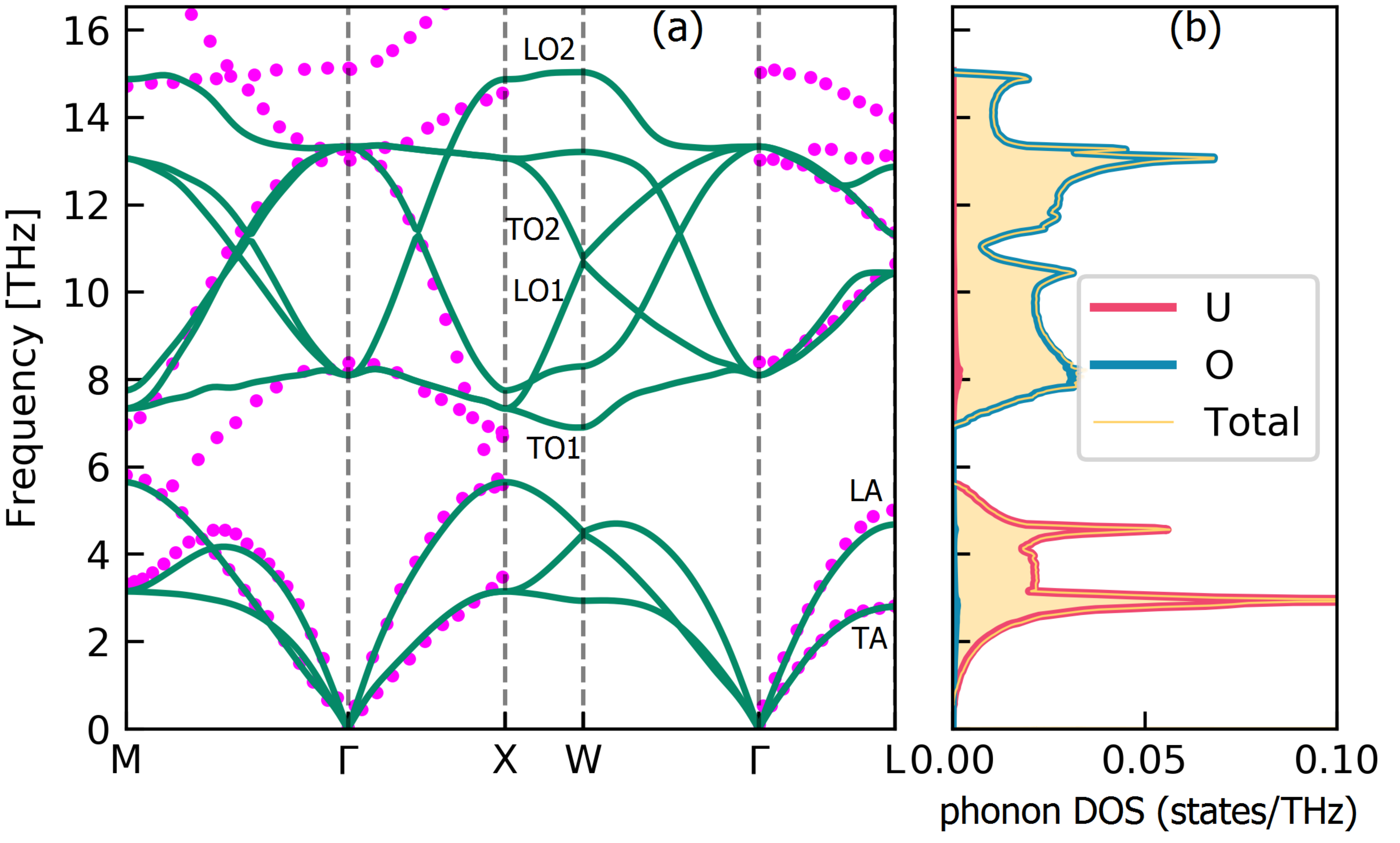}
	\caption{(a)- Phonon dispersion curves along the high-symmetry directions in BZ with solid violet balls corresponding to experiment\cite{dolling1965crystal}, and (b)- Phonon density of states.     }
	\label{fig1}
\end{figure}

Using the optimized geometry lattice parameter, the 2nd-IFCs were calculated employing the DFPT method. 
To this end, we first performed an scf calculation using the optimized lattice parameter with kinetic 
energy cutoffs of 60 and 600~Ry for the wavefunction and density expansion in terms of plane waves with a 
$\Gamma$-centered $8\times 8\times 8$ mesh in $k$ space with a tighter convergency 
threshold of $10^{-12}$~Ry. Using the resulting wavefunctions and eigenvalues as unperturbed quantities, 
the inhomogeneous KS equations in DFPT were solved in $q$-space self-consistently for the potential and 
density variations and thereafter the dynamical matrix is calculated using the atomic mass of U-238 isotope. The threshold for self-consistency 
of the potential variation was taken as $10^{-14}$. Diagonalizing the dynamical matrix, the phonon 
frequencies were obtained. In Fig.~\ref{fig1} the experimental and the calculated phonon dispersion curves 
along the high-symmetry directions in the BZ as well as the PhDOS are shown.

In Fig.~\ref{fig1}(a), among the nine calculated branches, the lowest three belong to transverse acoustic (TA) and longitudinal 
acoustic (LA) that correspond to the vibrations of heavy U atoms. The higher frequency branches belong to 
the transverse optical (TO) and longitudinal optical (LO) modes that correspond to vibrations of lighter O 
atoms. As is seen, the agreement between experiment and our calculation is excellent for the acoustic 
branches. However, because we have not included the non-analytical terms in the dynamical matrices in our  
simplified model, near the $\Gamma$ point in the $M-\Gamma -X$ and $\Gamma -L$ paths we see no splittings 
of LO-TO branches. As we will see below, these fine structures does not spoil the calculated lattice thermal conductivity 
which is a sum over all branches. 

In Fig.~\ref{fig1}(b), the plotted PhDOS shows that the states in 
acoustic branches belong to the vibrations of the heavy U atoms whereas the states of optical branches are 
comprised from the vibrations of the lighter O atoms.
It should be mentioned that the calculated phonon frequencies using the atomic mass of U-235 isotope did 
not result in a meaningful differences in the phonon dispersion, and therefore, we use the same phonon 
frequencies obtained from U-238 in the calculations of thermal conductivities of bulk UO$_2$ with 
different isotope abundances.

\begin{figure}
	\centering
	\includegraphics[width=\linewidth]{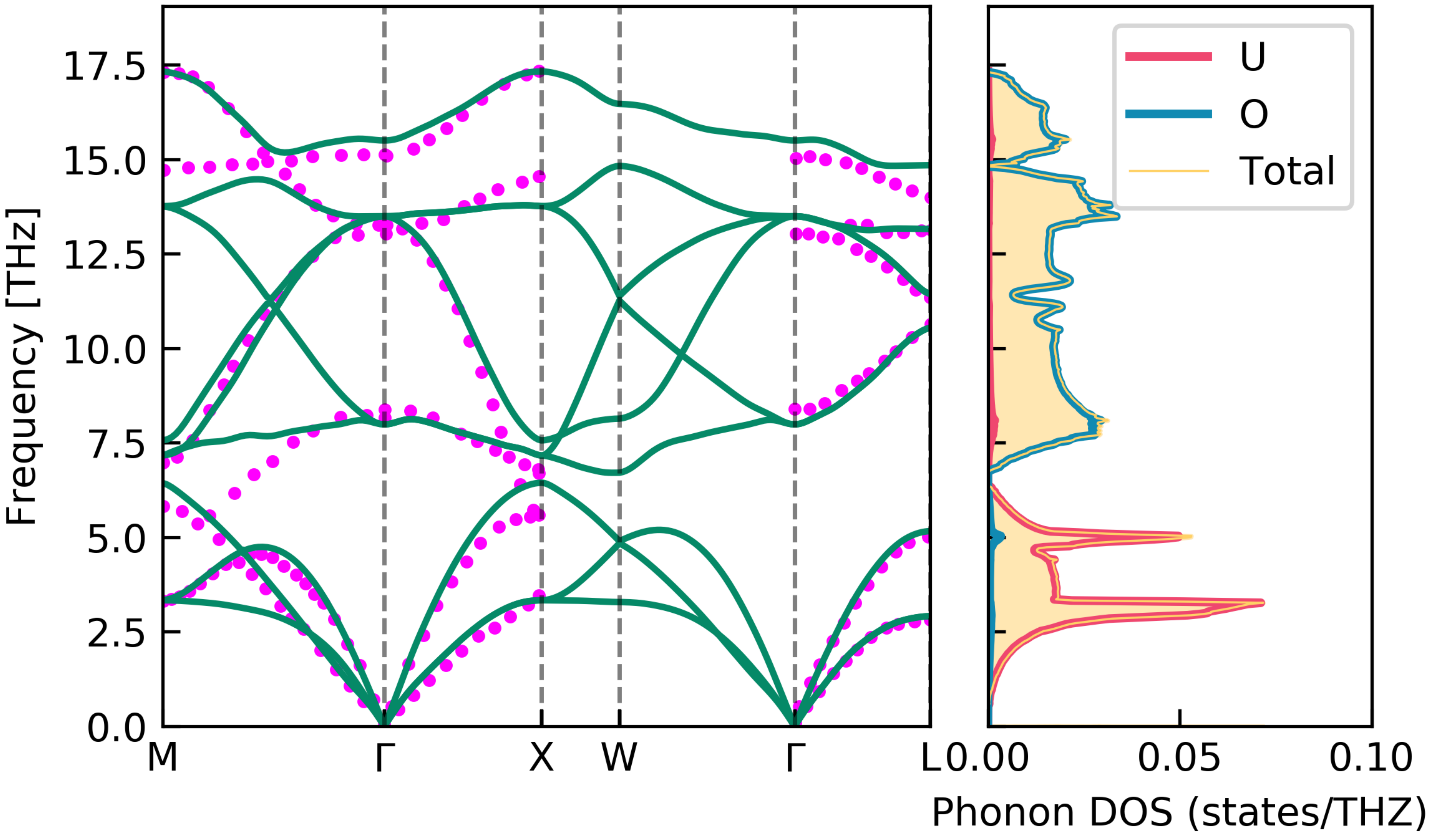}
	\caption{The same as in Fig~\ref{fig1}. Here, because of applying the non-analytical term correction, the LO-TO splitting is reproduced around the $\Gamma$ point, giving an excellent agreement with experiment.  }
	\label{fig2}  
\end{figure}

Before starting the discussion on thermal conductivity, as was mentioned earlier, in reality the bulk UO$_2$ solid is
electronically an insulator\cite{SHEYKHI201893} with an experimental gap of 2.1~eV. To obtain the exact phonon dispersion curve, some people\cite{pang2013} have used the sophisticated DFT+U method. However, in our simple model, the "exact dispersion" showing LO-TO splitting is obtained just by applying the non-analytic correction to the dispersion of Fig~\ref{fig1}(a). In the correction, the dielectric constant was chosen as 5.2, and the Born charges of U and O as 4.7 and -2.35 units.
The result is shown in Fig~\ref{fig2}. As is seen from Fig~\ref{fig2}(a), the non-analytic term correction applied on the second-order force constants in our simple model, could reproduce the LO-TO splitting around the $\Gamma$ point which leads to excellent agreement with experimental results.

\subsection{Thermal conductivity}
\begin{figure}
	\centering
	\includegraphics[width=0.8\linewidth]{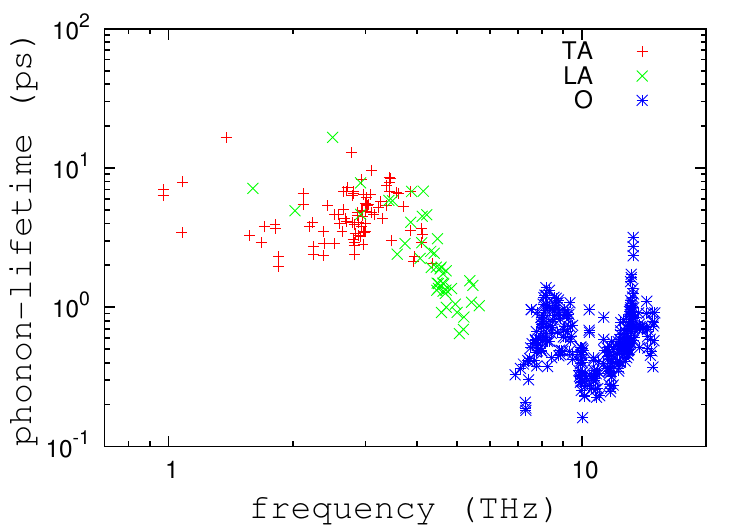}
	\caption{Phonon lifetimes in ps at $300^\circ K$. The red, green, and blue symbols correspond to TA, LA, and optical branches, respectively.    }
	\label{fig3}
\end{figure}

From Eq.(\ref{kappa}),to calculate the thermal conductivity we need phonon frequencies, equilibrium distribution function, group velocities, and phonon lifetimes. All needed quantities but phonon lifetimes were determined from our harmonic calculations. To calculate the phonon lifetimes we have solved the equation Eq.(\ref{tau}) by iteration until self-consistency. The calculated phonon lifetimes at $300^\circ K$ are shown in Fig.~\ref{fig3}.  
As seen from Fig.~\ref{fig3}, the TA branches have the largest lifetimes and therefore have dominant role in thermal conductivity. On the other hand, we expect that the optical branches having the shortest lifetimes (corresponding to the vibrations of O atoms) to have smaller contributions in the total conductivity. This result contradicts the result reported in Ref.\cite{pang2013} which claims the optical phonons have the largest contributions in thermal conductivity. However, the experimental results reported in Ref.~\cite{pang2013} that show the dominant role is played by optical phonons should be verified in another theoretical study using DFT+U or other methods.

\begin{figure}
	\centering
	\includegraphics[width=0.8\linewidth]{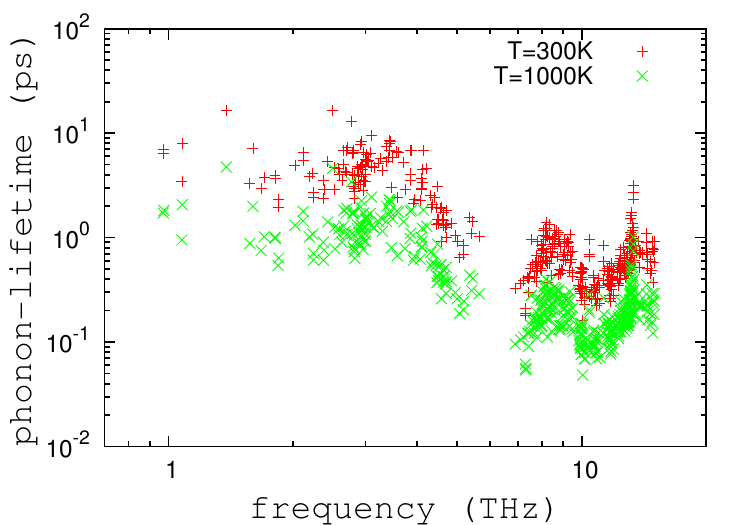}
	\caption{Calculated phonon lifetimes in ps. The red and green symbols correspond to $300^\circ K$ and $1000^\circ K$, respectively.}
	\label{fig4}
\end{figure}

In Fig.~\ref{fig4}, we have shown the lifetimes of the phonons at two temperatures of $300$ and $1000^\circ K$. It is evidently seen that the phonon lifetimes decrease with temperature.       

\begin{figure}
	\centering
	\includegraphics[width=0.8\linewidth]{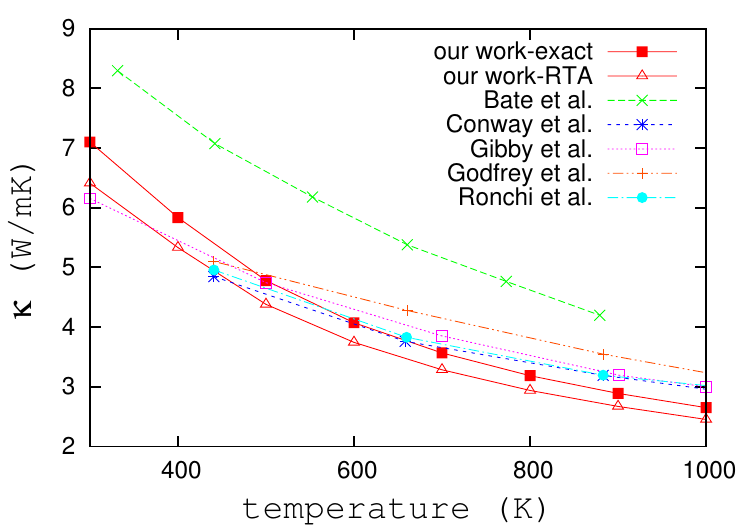}
	\caption{Lattice thermal conductivity of UO$_2$, in W/mK, as a function of temperature. The red solid squares and red open triangles correspond to this work with full-iterative method and RTA calculations, respectively which are compared with experimental data.}
	\label{fig5}
\end{figure}

In Fig.~\ref{fig5}, the calculated thermal conductivities in both RTA and full-iterative schemes are 
compared with experimental data. As is seen, the agreement between full-iterative results and 
experiment is very good at relatively lower temperatures. This is because, in experiment at lower temperatures the four- 
and higher- phonon processes are not activated and the dominant contribution comes from the three-phonon 
processes which is consistent with our calculations. 

\begin{figure}
	\centering
	\includegraphics[width=0.8\linewidth]{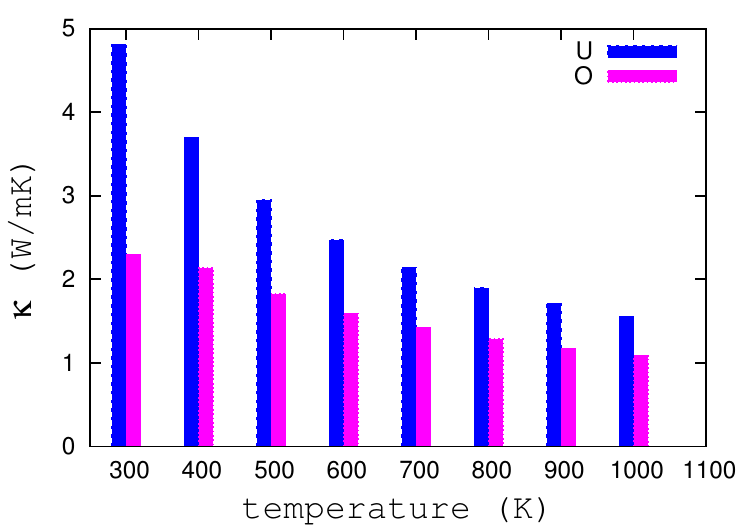}
	\caption{Elemental resolution of total thermal conductivity. Blue and violet bars correspond to U and O contributions, respectively.}
	\label{fig6}
\end{figure}

In Fig.~\ref{fig6}, the thermal conductivity is resolved into the elemental contributions and as is seen, the contributions from U atoms are dominant at all temperatures. This is consistent with the PhDOS plot of Fig.~\ref{fig1}(b) in which all contributions of acoustic branches comes from the U atoms.  

\begin{figure}
	\centering
	\includegraphics[width=0.8\linewidth]{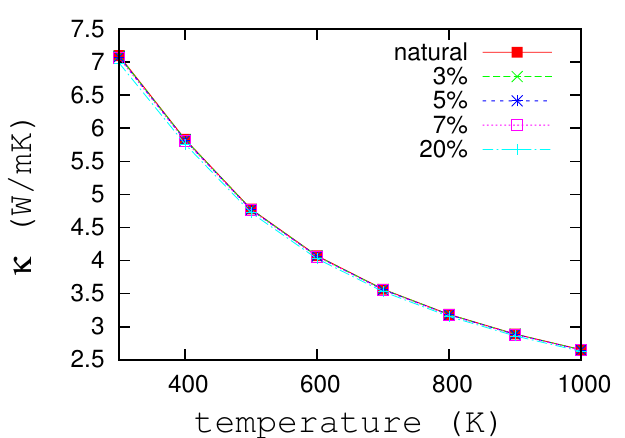}
	\caption{Thermal conductivities in W/mK as  functions of temperatures for fuels with different abundances of U-235. }
	\label{fig7}
\end{figure}

\begin{figure}
	\centering
	\includegraphics[width=0.8\linewidth]{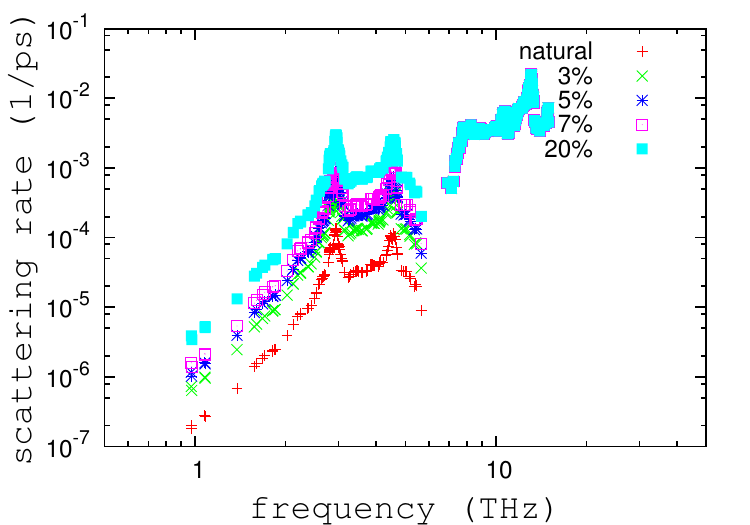}
	\caption{Phonon scattering rate due to isotopic mass disorder in ps$^{-1}$ with respect to frequencies for different U-235 concentrations. The lower scattering rates correspond to lower U-235. }
	\label{fig8}
\end{figure}

Finally, in Fig.~\ref{fig7}, we have plotted the thermal conductivities for fuels with different abundances of U-235 isotope. 
Inspecting the values for different abundances, we observe that with increasing the concentration of U-235, the thermal conductivity decreases, which is due to the increasing the scattering rates due to impurities, $\Gamma_{\lam\lamp}$. This argument is confirmed by looking at Fig.~\ref{fig8}. However, the decrease of thermal conductivity is not significant because, the phonon lifetime, which is determined from the combination of anharmonic and isotopic scattering rates through Matthiessen's rule\cite{matthiessen1864iv}, does not change significantly.

\section{Conclusions}\label{sec4}
In this work the lattice thermal conductivity of bulk UO$_2$ is studied by the exact solution of the BTE 
equation for the steady-state phonon distribution function. 
In this lattice-thermal-conductivity study, we have modeled the high-$T$ paramagnetic state of bulk UO$_2$ 
by a non-spin-polarized calculation, and neglected the Hubbard-U correction for the localized f electrons 
in the U atoms.
The computed phonon spectra showed that among the nine branches, the lowest three belonging
TA and LA, corresponding to U atoms,
were in excellent agreement with experimental data. However, because in the simplified model we
had not included the non-analytical terms in the dynamical matrices, no LO-TO splittings were observed  
near the $\Gamma$ point in the $M- \Gamma -X$ and $\Gamma -L$ paths in phonon spectra. Even with 
existing these differences in the experimental phonon spectra and our calculated results, the computed thermal 
conductivity was in good agreement with experiment because, the sum over phonon branches in 
Eq.(\ref{kappa}) somehow washes out the detailed information of the phonon spectra.    
Our calculated thermal conductivity showed a small deviation from the experiments which was due to the 
fact that we had only taken into account the three-phonon processes while in reality the higher-order 
heat-carrying processes come into play at higher temperatures. 
The calculated phonon lifetimes at 300$^\circ K$ showed that the TA branches had the
largest lifetimes and therefore had dominant role in thermal conductivity. On
the other hand, the optical branches having the shortest lifetimes (corresponding to the vibrations of O 
atoms) had smaller contributions in the total conductivity. 
Due to the fact that nuclear power reactors may use fuels with different relative concentrations of U-235 
and U-238, we had also repeated the thermal-conductivity calculations for the cases with U-235 
concentrations 
of 3\%, 5\%, 7\%, and 20\%, and observed a small decrease of thermal conductivity by increasing the 
impurity level. This fact was explained to be due to the increase of the phonon scattering rates from the 
impurity atoms (U-235).  

\section*{Acknowledgement}
This work is part of research program in School of Physics and Accelerators, NSTRI, AEOI.

\appendix

\section{Convergency tests}\label{app}
In the course of thermal conductivity calculations, at each step we have to ensure the convergency 
of the calculated quantities with respect to relevant parameters. These steps include the DFT 
calculations, harmonic lattice-dynamic calculations, 3rd-IFCs calculations, and the solution of the 
BTE. First we ensure the convergency of lattice constant with 
respect to basis-set kinetic-energy cutoffs and $k$-grid for structural optimization. The results are 
tabulated in Table~\ref{tab2}:

\begin{table}[]\small
	\caption{Zero-pressure lattice constant, in $\AA$, as a function of $k$-mesh and plane-wave energy cutoff, in Ry, for wavefunction expansion. The cutoff for density expansion is taken as 10 times of $E_c$. The chosen parameters for $k$-mesh and $E_c$ are $8\times 8\times 8$ and 60~Ry, respectively; and the corresponding lattice constant is 5.3968$\AA$.}
	\centering
	\small
	{\begin{tabular}{cccccc} \hline\hline
			&      &      &$E_c$  &     &   \\ \cline{2-6}
			$k$-mesh        & 40   & 50   & 60    & 70  & 80      \\ \hline
			$2\times 2\times 2$ & 5.4007& 5.4011& 5.4010& 5.4014& 5.4015    \\ 
			$3\times 3\times 3$ & 5.3993& 5.3994& 5.3996& 5.3999& 5.3999    \\ 
			$4\times 4\times 4$ & 5.3937& 5.3937& 5.3939& 5.3943& 5.3942    \\ 
			$5\times 5\times 5$ & 5.3979& 5.3985& 5.3982& 5.3987& 5.3987    \\ 
			$6\times 6\times 6$ & 5.3956& 5.3957& 5.3959& 5.3962& 5.3965    \\ 
			$7\times 7\times 7$ & 5.3965& 5.3965& 5.3963& 5.3967& 5.3966    \\ 
			$8\times 8\times 8$ & 5.3965& 5.3968& {\bf 5.3968}& 5.3970& 5.3969    \\
			$10\times 10\times 10$ & 5.3965& 5.3963& 5.3964& 5.3971& 5.3966    \\ \hline\hline
		\end{tabular}}
	\label{tab2}
\end{table}

We have chosen a $\Gamma$-centered $8\times 8\times 8$, 60~Ry, and 600~Ry for the $k$-mesh, wavefunction cutoff, and density cutoff, respectively for our DFT calculations and obtained the converged value of $5.3968\AA$ for lattice constant.

In the DFPT calculations, we have started with a $q$-grid of $8\times 8\times 8$ and calculated the phonon 
frequencies and the 2nd-IFCs. 
To calculate the 3rd-IFCs, we have used the supercell approach. 
In this step we determine the appropriate supercell size, magnitude of atomic displacement, and the number 
of neighboring shells.   
The appropriate supercell size is intimately connected to the range of forces\cite{mcgaughey2019phonon}. 
We have displaced one atom 
in a $3\times 3\times 3$ supercell with 81 atoms from its equilibrium position by 0.2~Bohr. Then, we 
determined the magnitude of forces acting on each of 81 atoms in the supercell and plotted with respect to 
the distance from the displaced atom (See Fig.~\ref{fig8}).  
\begin{figure}
	\centering
	\includegraphics[width=0.8\linewidth]{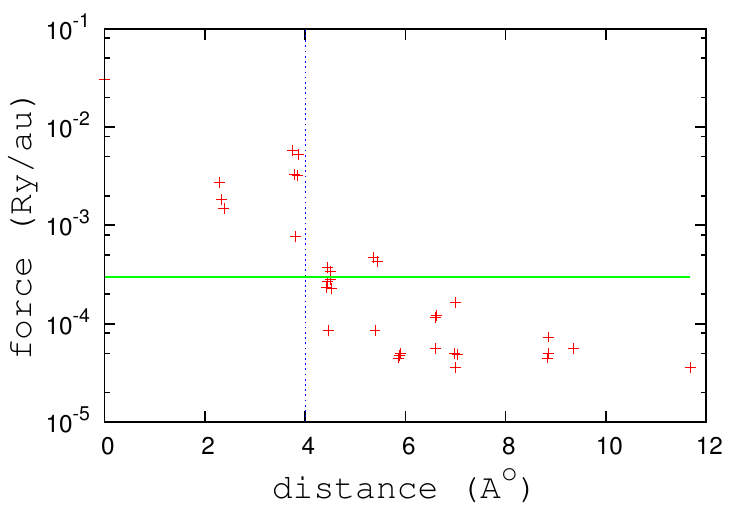}
	\caption{Force, in Ry/au, on each atom of the 81-atom supercell with one atom displaced from equilibrium position. The horizontal green line represents the 1\% of the force on displaced atom. The vertical dashed blue line specifies the starting cutoff for calculating 3rd-IFCs.}
	\label{fig9}
\end{figure}
As shown in Fig.~\ref{fig9}, the forces drop to $\sim 0.01$ of the maximum force at a distance of $4\AA$  
which encloses up to second neighbor atomic shells. This estimate indicates that the adopted supercell size of $3\times 3\times 3$ is appropriate for our 3rd-IFCs calculations. 
Using this force cutoff as an initial guess for the cubic cutoff, we have then determined the appropriate 
magnitude for the atomic displacements. To this end, considering all the symmetries of the 
system, we have generated sets of 108 displacements sufficient for calculating 3rd-IFCs for each magnitude 
of 0.005, 0.01, 0.03, 0.05, and 0.1$\AA$. Testing the thermal conductivities obtained from these trial 
3rd-IFCs (See Fig.~\ref{fig10}), we found that the appropriate magnitude of $0.03\AA$ gives stable value for the thermal 
conductivity at $300^\circ K$. 

\begin{figure}
	\centering
	\includegraphics[width=0.8\linewidth]{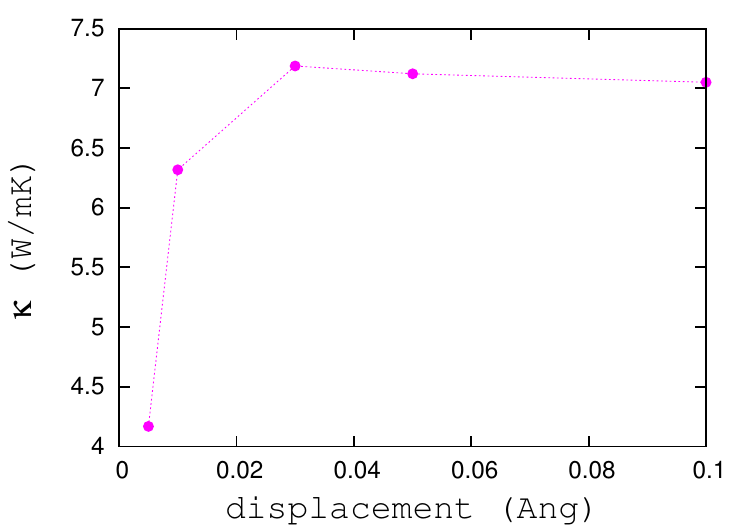}
	\caption{Lattice thermal conductivity of UO$_2$ at $T=300^\circ K$ as a function of atomic displacement size. The value of $0.03\AA$ is chosen for 3rd-IFCs calculations.}
	\label{fig10}
\end{figure}

Next, we have tested the convergency of thermal conductivity with respect to the number of interaction shells in 3rd-IFCs calculation. Due to significant increasing of the number of displaced structures (108 compared to 208), but no significant gain in accuracy of thermal conductivity, we adopted up to second neighboring shell in our calculations (See Table~\ref{tab3}). 

\begin{table}[]\small
	\caption{Cubic cutoff [\# of shells/distance ($\AA$)] in terms of number of neighboring shells/distance, the resulting number of displaced structures, and the computed thermal conductivity.     }
	\centering
	{\begin{tabular}{ccc} \hline\hline
			cubic cutoff  &\;\;\;  \# of disp. struct.  & \;\;\;\;$\kappa$ (W/mK)  \\ \hline\hline 
			1/3.5             & 44       & 7.8728       \\ 
			{\bf 2}/{\bf 4.0} & {\bf 108}& {\bf 7.2686} \\
			3/5.0             & 176      & 7.1736       \\
			4/5.4             & 208      & 7.2347       \\ \hline\hline
		\end{tabular}}
	\label{tab3}
\end{table}

Finally, we have tested the convergency of thermal conductivity with respect to the $q$-mesh in the DFPT calculations, and found that the $q$-mesh of $12\times 12\times 12$ leads to converged result.  

\section*{Data availability }
The raw or processed data required to reproduce these results can be shared with anybody interested upon 
sending an email to M. Payami.

\bibliography{payamisheykhi-revised-arxiv-99.02.09}

\end{document}